\documentclass[a4paper,11pt,oneside]{article}
\usepackage[english]{babel}
\usepackage{graphicx}
\usepackage[svgnames]{xcolor}
 
\usepackage{amssymb}
\usepackage[tbtags]{amsmath}

\usepackage[T1]{fontenc}
\usepackage{lmodern}

\usepackage[top=32mm]{geometry}

\usepackage{caption}
\usepackage[authoryear,round,longnamesfirst]{natbib}
\newcommand{\tightlist}{%
  \setlength{\parskip}{0pt}\setlength{\itemsep}{0pt}}
		
\usepackage{fancyhdr}
\usepackage{hyperref}
\hypersetup{
  linktocpage=true,
  linkcolor  = MidnightBlue,
  citecolor  = MidnightBlue,
  urlcolor   = MidnightBlue,
  colorlinks = true,
}

\usepackage{booktabs}

\title{Subexponential growth of early Christianity}
\author{Jorge C. Lucero\thanks{Dept.\ Computer Science, University of Bras\'{i}lia, Brazil. E-mail: \href{mailto:lucero@unb.br}{lucero@unb.br} }} 
\date{\today}

\begin{document}

\maketitle

\begin{abstract}
This paper presents a simple mathematical model for the growth of the Christian population in the Roman Empire during the first to fourth centuries. The model has a subexponential growth rate of order $e^{o(t)}$, where $o$ denotes the ``little-o'' asymptotic bound, but still superpolynomial, and it fits available Christian population estimates with good accuracy.
\end{abstract}

\section{Introduction}

\label{intro}

It is generally accepted that, in its beginning, Christianity grew from a few individuals at the start of the first century to about half the population of the Roman Empire at the end of the fourth century \citep{Ehrman2018,Stark1997}. A first mathematical description of such a dramatic growth was introduced by \citet{Stark1997}. In his work, Stark estimated a Christian population of 1000 individuals in year 40 CE, and around 10\% of a total Empire population of 60 million in year 300 CE. Then, he showed that those numbers could be produced with an exponential function and a growth rate of 40\% per decade. 

More recently, \citet{Ehrman2018} contested \citeauthor{Stark1997}'s \citeyearpar{Stark1997} numbers. Ehrman assumed around 20 Christians in year 30 CE, time of Jesus' death, and claimed that Stark's estimate of 1000 Christians just a decade later would be too high. Further, a growth rate of 40\% per decade would imply in 170 million Christians in the Roman Empire by year 400, surpassing its total population. Then, Ehrman modified Stark's model by still assuming exponential growth, but with different growth rates at different time periods: 300\% from 30 CE to 60 CE, 60\% from 60 CE to 100 CE, 34\% during the second and third centuries, and 26\% during the fourth century. This model produced more realistic values; e.g., only 80 Christians in 40 CE (against Starks's 1000 estimate),
and almost 30 million Christians in 400 CE, matching the accepted estimate of half the Empire's total population.  Further, the model produced plausible results at intermediate years; e.g., 1280 Christians at 60 CE, when a number of Christian churches had already been founded in the eastern Mediterranean region, and around three million Christians at the beginning of the fourth century and time of Emperor Constantine conversion to Christianity. 

\citeauthor{Ehrman2018}'s \citeyearpar{Ehrman2018} model is a piecewise exponential one with a decreasing growth rate, which indicates an overall subexponential growth for the whole period from 30 CE to 400 CE. Here, we consider a function $f(t)$ to have subexponential growth if it increases at slower rate than any exponential function $b^{t}$, with $b>1$. Subexponential growth patterns have been observed in early stages of epidemics of, e.g., COVID-19, HIV, Ebola, and other diseases \citep{Chowell2016a, Maier2020, Viboud2016}, instead of the expected exponential one, and they have been attributed to spatial heterogeneity, clustering of contacts, and reactive behavioral changes of the population \citep{Chowell2016}. 
In fact, the spread of religious beliefs share similarities with the spread of infectious diseases, and epidemiological models have been applied to characterize modern church growth and decline \citep{Bettencourt2006,Hayward2005,Jo2021,McCartney2015}. It may be then interesting to explore how well subexponential functions can fit the early Christianity data, which is the purpose of this paper.

We also must note that, in recent years, a variety of more sophisticated modeling approaches have been applied to describe church growth, decay and competition, such as compartmental models \citep{Bettencourt2006,Hayward2005,Jo2021,McCartney2015}, stochastic models \citep{Picoli2008}, and Verhulst-Lotka-Volterra's models \citep{Vitanov2010}. However, those models would present difficulties to the present study, given the sparsity of reliable data available (at the most, rough estimates of four data points). Therefore, we will restrict our study to simple continuous functions with no more than three fitting parameters.

\section{Data}   
\label{data}
We consider \citeauthor{Ehrman2018}'s \citeyearpar{Ehrman2018} estimates:
\begin{enumerate}
\tightlist
\item 20 Christians in year 30 CE, at the estimated time of Jesus' death.
\item 1000 -- 1500 Christians in year 60 CE, at the estimated time of the last Pauline epistle in the New Testament.
\item 2 -- 3 million Christians in the year 300 CE, equivalent to 3.5 -- 5\% of the total population of the Roman Empire estimated at 60 million people.
\item 30 million Christians in the year 400 CE, equivalent to half the total population of the Roman Empire.
\end{enumerate}

\section{Mathematical models}

We adopt a definition from the theory of computational complexity, which states that a function $f(t)$ grows subexponentially in time if $f(t)=e^{o(t)}$, where $o$ is the ``little-o'' asymptotic bound \citep{Kaliski2011}. This definition means that $f(t)$ is asymptotically smaller than any function of the form $e^{kt}$, where $k$ is a positive constant,\footnote{The base of the exponential is irrelevant here, since any function $f(t)=b^{\ell t}$ may be written as $f(t)=e^{(\ell\ln b)t}$.} and it implies \citep{Sipser2012}
\begin{equation}
\lim_{t\rightarrow\infty} \frac{\ln f(t)}{t}=0.
\label{def1}
\end{equation}

Letting $y=f(t)$ be the Christian population at a time $t$, we may satisfy Eq.~\eqref{def1} by setting 
\begin{equation}
\ln y = a\sqrt{t-b} + c, 
\label{mod1}
\end{equation}
where $a>0$, $b\le 30$ and $c>0$ are real coefficients.\footnote{Subexponential growth may be also defined as order $O(e^{t^\varepsilon})$, for every $\varepsilon > 0$, where $O$ is the ``big-O'' asymptotic bound \citep{Wegener2005}. This definition is not satisfied by the proposed model for $0< \varepsilon <1/2$. In this context, the model would be considered still as exponential, since it has a lower asymptotic bound of $e^{t^\varepsilon}$ for some $\varepsilon > 0$ (e.g., $\varepsilon=1/2$), although not ``strictly'' exponential.} Using the data in Section \ref{data} and a least square optimization algorithm, we obtain $a=0.731$, $b=30$ and $c = 3.008$. 

\begin{figure}
\centering
\includegraphics{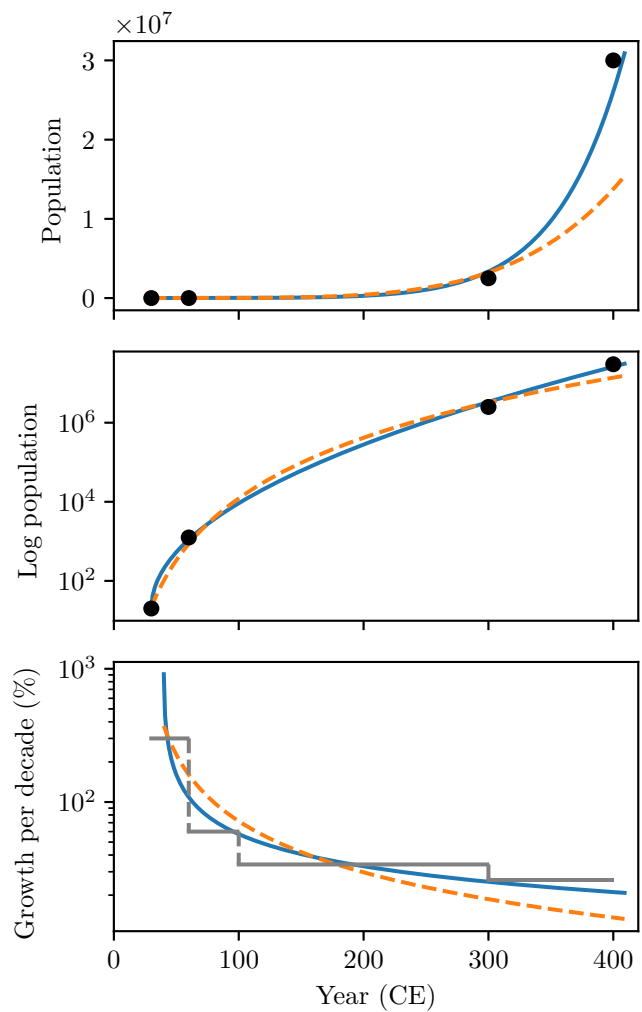}
\caption{Estimates and modeled results of early Christian population. Top: population. Middle: log population. Bottom: relative growth rate per decade. Solid blue curve: results from Eq.~\eqref{mod1}. Dashed orange curve: results from Eq.~\eqref{mod2}. Black dots (top and middle plots) and gray ``staircase'' (bottom plot): \citeauthor{Ehrman2018}'s \citeyearpar{Ehrman2018} estimates.}
\label{fig1}
\end{figure}

The predictions of the model are shown in Table \ref{table1} and Fig.~\ref{fig1}. In general, it provides a close fit to the data, with an rms relative error of 19\%. The bottom plot in Fig.~\ref{fig1} shows the relative growth rate $r(t)$ per decade, computed as
\begin{equation}
r(t) = \frac{y(t) - y(t-10)}{y(t-10)}\times 100,
\label{rt}
\end{equation}   
where $t$ is measured in years. The growth rate decreases monotonically from 910\% at year 40 CE to 21\% at year 400 CE. The plot also shows the growth rates given by \citet{Ehrman2018}, in the shape of a decreasing ``staircase'' where each horizontal portion corresponds to a period of exponential growth. This ``staircase'' is also well approximated by the model.

\begin{table}
\centering

\caption{Estimates \citep{Ehrman2018} and modeled results of early Christian population.}
\label{table1}

\begin{tabular}{rrrr}
\toprule
Year (CE) & Estimate  & Eq.~\eqref{mod1} & Eq.~\eqref{mod2}\\
\midrule
30 & 20 & 20 & 22\\
60 &  1000 -- 1500 & 1112 & 867 \\
300 & 2 -- 3 million & 3.4 million & 3.2 million\\
400 & 30 million & 26.1 million & 13.8 million\\
\bottomrule
\end{tabular}

\end{table}

Eq.~\eqref{mod1} is subexponential but still it grows faster than a polynomial function. This fact may be assessed by computing the limit
\begin{equation}
\lim_{t\rightarrow\infty} \frac{y(t)}{t^n}=\infty, \quad \text{for any integer $n$}.
\label{def2}
\end{equation}

For comparison, let us consider a second model which comes from the field of epidemiology \citep{Viboud2016}, and is defined by the differential equation
\begin{equation}
y'=ay^p,
\end{equation}
where $a$ is a positive real coefficient and $p=1-(1/b)$, where $b$ is a positive integer. Its general solution is the polynomial  
\begin{equation}
y(t) = \left(\frac{a}{b}t+c\right)^b,
\label{mod2}
\end{equation}
where $c$ is an integration constant \citep{Tolle2003}.

Again, using the data in Section \ref{data} and a least square optimization algorithm, we obtain $a=0.336$, $b = 5$ and $c=-0.173$.  The results are also shown in Table \ref{table1} and Fig.~\ref{fig1}. The rms relative error is 36\%, indicating a poorer fit than the previous model. Particularly, the top plot in Fig.~\ref{fig1} shows that the polynomial is not able to achieve the same growth pattern of the data.

\section{Conclusion}

The dramatic growth of the Christian population in the first to fourth centuries seems to follow a subexponential pattern; i.e., slower than an exponential but still faster than a polynomial one. As a next step, it might be interesting to investigate if similar patterns apply to other religions at their beginnings.

Another interesting issue concerns the conversion of the Roman Emperor Constantine in year 312 CE. In his book, \citet{Ehrman2018} noted that although Constantine's conversion may have contributed to the growth of Christianity in the empire, the conversion does not seem to have produced a noticeable impact in the growth pattern. The present results support such a claim, since a single subexponential function is able to take Christianity from a few individuals to several millions in the lapse of four centuries, without the need to introduce any additional factor at the start of the fourth century.

\end{document}